\documentstyle{article}
\title{The Gravitational Energy of a Point Mass is Finite}

\author{L.V.Verozub}

\begin{document}

\maketitle
\centerline{Kharkov State University, Kharkov 310077, Ukraine}
\centerline{ (verozub@gravit.kharkov.ua)}

\begin{abstract}
We argue that  our gravitation equations \cite{Verozub1}
in the flat space-time lead to the finite proper gravitational energy
of a point mass .
\end{abstract}

\section{The Gravitation Equations}
Perhaps \cite{Thirring} , gravitation can be described by a tensor
field $\psi _{\mu \nu }$  in flat space-time
and the Lagrangian action for the test point masses
$m$  in this field is of the form
\begin{equation}
L = -mc[g_{\alpha \beta }(\psi)\; \dot{x}^{\alpha }\;\dot{x}^
{\beta }]^{1/2} ,
\label{LagrangianThirr}
\end{equation}
where
$\dot{x}^{\alpha } = dx^{\alpha }/dt$ and $g_{\alpha \beta }$ is a
symmetric tensor whose components are the function of
$\psi_{\alpha \beta }$.

The field $\psi_{\mu \nu }$ in flat space-time is an analog  of the
potential $A_{\mu }$ of the electromagnetic field. Therefore the field
equations for $g_{\mu \nu }(\psi)$ must be invariant under the gauge
transformations
$\psi_{\mu \nu } \rightarrow \overline{\psi_{\mu \nu }}$>. The
simplest
equations of such a kind where proposed in paper \cite{Verozub1} .
The equations can be written in the form

\begin{equation}
     B_{\beta \gamma ,\alpha  }^{\alpha } - B_{\beta \mu }^{\nu } \;
     B_{\gamma \nu }^{\mu } = 0 \;,
\label{myeq2}
\end{equation}
where the comma denotes the covariant derivative with respect to the
metric
tensor $\eta_{\mu \nu }$ in pseudo-Euclidean  space-time $E$.

In this equations  the tensor
$B_{\beta \gamma }^{\alpha }$ is given by
\begin{equation}
      B^{\gamma }_{\alpha \beta } = \Pi ^{\gamma }_{\alpha \beta } -
       \Pi ^{\gamma }_{\alpha \beta }
\label{tensB}
\end{equation}
(Greek indices run from 0  to 3) , where
\begin{equation}
   \Pi ^{\gamma }_{\alpha \beta } =
\Gamma ^{\gamma }_{\alpha \beta } -
(n+1)^{-1} \left [
\delta ^{\gamma }_{\alpha } \Gamma ^{\epsilon }_{\epsilon \beta } -
\delta ^{\gamma }_{\beta } \Gamma ^{\epsilon }_{\epsilon \alpha }]
\right ] ,
\label{Thomases}
\end{equation}
\begin{equation}
\stackrel{\circ}{\Pi} ^{\gamma }_{\alpha \beta } =
\stackrel{\circ}{\Gamma} ^{\gamma }_{\alpha \beta } -
(n+1)^{-1}
\left [ \delta ^{\gamma }_{\alpha } \stackrel{\circ}{\Gamma }^
{\epsilon }_
{\epsilon \beta } -
\delta _{\beta }^{\gamma } \stackrel{\circ}{\Gamma }_{\epsilon
\alpha }
^{\epsilon }  \right ] ,
\label{Thomases0}
\end{equation}
$\stackrel{\circ}{\Gamma }_{\alpha \beta }^{\gamma }$
 are the Christoffel symbols in   $E$ and
$\Gamma _{\alpha \beta }^{\gamma }$    are the Christoffel symbols
of the
Riemannian space-time $V$ of dimenstion $n$ , whose  fundamental
tensor is
$g_{\alpha \beta }$ .

The tensor  $B_{\alpha \beta }^{\gamma }$ can be formed by replacing
the odinary derivatives in $\Pi _{\alpha \beta }^{\gamma }$ with
the covariant ones in $E$.

$B_{\beta \gamma ,\alpha  }^{\alpha }$ satisfies also the following
identities:

\begin{equation}
\stackrel{\ast}{R}_{\alpha \beta \delta }^{\gamma  }  +
\stackrel{\ast}{R}_{\beta \delta \alpha }^{\gamma  }  +
\stackrel{\ast}{R}_{\delta \alpha \beta }^{\gamma }   = 0   ,
\label{identity}
\end{equation}
where tensor
$\stackrel{\ast}{R}_{\alpha \beta \delta}^{\gamma  }$
is obtained from the Riemannian curvarure tensor  by  replacing the
Christoffel symbols with the Thomas symbols and  by replacing  ordinary
derivative with the covariant one in $E$.

The equations (\ref{myeq2})    and (\ref{identity})  are field equations
for the
tensor $B_{\beta \gamma }^{\alpha }$.

    Eqs.(\ref{myeq2}) are invariant under arbitrary
transformations of the tensor $g_{\alpha \beta }$   retaining invariant
the
equations of
motion of a test particle , i.e.geodesic lines in $V$ . In  other words,
eqs.(\ref{myeq2})
are the geodesic-invariant . Thus , the tensor field $g_{\alpha \beta }$
  is
defined  up
to   geodesic  mappings of space-time $V$ (in the way analogous  to
the   definding the potential  $A_{\mu }$   in  electrodynamics
up  to  gauge transformations).
Therefore, additional conditions can be imposed on the tensor
$g_{\alpha \beta }$   . In
partucular, if  the tensor  $g_{\alpha \beta }$   satisfies
the conditions
\begin{equation}
Q_{\alpha } = \Gamma _{\alpha \sigma }^{\sigma } -
 \stackrel{\circ}{\Gamma} _{\alpha \sigma }^{\sigma } = 0 ,
\label{gaugeConditions}
\end{equation}
then eqs (\ref{myeq2}) will be  reduced  to vacuum Einstein equations
$R_{\alpha \beta } = 0$ ,
where $R_{\alpha \beta }$   is the Richi tensor \cite{Verozub1}.
    Unlike the $g_{\alpha \beta }$    or
$\Gamma _{\alpha \beta }^{\gamma }$    the tensor
$B_{\alpha \beta }^{\gamma }$   is invariant
under the geodesic mappings  of space-time $V$ as well as
strength tensor $F_{\alpha \beta }$ in electrodynamics  is invariant
  under the gauge transformations.

\section{Spherically Symmetric Field}
For a spherically symmetric field the nonzero components of the tensor
$B_{\alpha \beta }^{\gamma }$ in the spherical coordinate system are:

\begin{center}

$ B_{11}^{1}=A'/2A ,\; \; B_{22}^{1}= -r(A^{-1}-1),\; \;
B_{10}^{0}=C'/2C$, \\

\vspace{5mm}

$B_{33}^{1} = -r \sin ^{2}(\theta) (A -1) ,\;\;
B_{00}^{1} = CA'/2A $ ,

\end{center}

where
the functions $A$ , $B$ and $C$ are :
\begin{equation}
A = (f')^2 (1-\beta /f)^{-1} ,\; B = f^2 ,\; C = 1 - \beta  /f
\label{ABCequal}
\end{equation}
and
\begin{displaymath}
f = ( r^{3} + \beta ^{3} )^{1/3}
\end{displaymath}
and $ f' = df / dr $ .  The constant $\beta $ is not determined from
the
boundary conditions \cite{Verozub1}. If we put $\beta =\alpha $  ,
where $\alpha =2GM/c^{2}$ , $M$ is the attractive mass, $G$ is the
gravitational constant and  $c$ is the speed of light, then  the
solution
(\ref{ABCequal}) has no a physical singularity in the sense that
this solution does not lead to the collapse. Some astrophysical
consequences of this solution are considered in \cite{Verozub2} .

\section{The Gravitational Energy of a Point Mass }
Eq. (\ref{myeq2}) without the nonlinear term is analogous to the
electrodynamics equations  $F_{\beta , \gamma }^{\gamma }= 0$ for the
strength tensor $F_{\alpha \beta }$.
Since the gravitational field must be self-interacting we can suppose
that eq. (\ref{myeq2}) are of the form
\begin{equation}
B_{\alpha \beta , \gamma  }^{\gamma }  = \kappa t_{\alpha \beta },
\label{B=t}
\end{equation}

where $ \kappa  = 4 \pi C /c^4 $ and

\begin{equation}
t_{\alpha \beta } = \kappa ^{-1}
B_{\alpha \gamma }^{\delta } \; B_{\beta \delta }^{\gamma }
\label{t=}
\end{equation}

is the
energy-momentum tensor of gravitation field   .

The components of the 3-momentum density vector
\begin{equation}
P_{i}= t_{0 \beta}  \; \; ( i = 1,2,3 )
\label{3-momentum}
\end{equation}
 are equal to zero
for the solution (\ref{ABCequal}).

Let us find the  energy of a point
masses
\begin{equation}
{\cal E} = \int t_{0}^{0} dV
\label{energydef}
\end{equation}

We have
\begin{equation}
t_{00}=2 \kappa ^{-1} B_{00}^{1} \; B_{01}^{0}
\label{t00}
\end{equation}
and ,therefore, in the used coordinates system

\begin{equation}
  {\cal E} = \int t_{0}^{0} dV = - \frac{1}{8} \frac{\alpha ^2 c^{2}}
{\pi \gamma } J ,
\label{energycalc}
\end{equation}
where

\begin{equation}
 J = \int \frac{dV}{f^{4}} = \frac{4 \pi}{3 \beta } B(1, 1/3) =
\frac{4 \pi}{\beta } ,
\label{J}
\end{equation}
where
\begin{equation}
B(z,w)= \int_{0}^{\infty} \frac{t^{z-1}}{ (1 + t)^{z+w}} dt
\label{B-function}
\end{equation}
is the B-function \cite{Abramovitz} . Using the equation
\begin{equation}
 B(z,w)= \frac{\Gamma (z) \Gamma (w)}{\Gamma (z+w)},
\label{BataGamma}
\end{equation}
where $\Gamma $ is $\Gamma $-function and setting   $\beta =\alpha $ ,
we obtain finally
\begin{equation}
{\cal E} = M \, c^{2}   .
\label{energyfinally}
\end{equation}
On the contrary, if we seach the constant $\beta $ from the
condition that gravitational energy of a point mass is given by
equation (\ref{energyfinally}) , then we shall arrive at the
conclution that $\beta =\alpha $.

Certainly , the 4-divergency of the tensor $t_{\alpha \beta }$ must be
equal to zero. It is likely that the 4-divergency of the tensor
$B_{\alpha \beta }^{\gamma }$ is not equal to zero identically.
However , it is should be observed that the conservation law for the
energy-momentum tensor must not be is satisfied for an arbitrary field
$B_{\alpha \beta }^{\gamma }$ (or $g_{\alpha \beta }$),
and only for the fields wich are solutions of  eq. (\ref{myeq2}).
For the received solution of the field equations  the
4-divergency of the tensor
$B_{\alpha \beta }^{\gamma }$ is indeed equal to zero. Thererfore,
the tensor $t_{\alpha \beta }$ for the spherical-symmetric field
is a conservation value.

The gravitation energy of the point mass being finite, it is
sufficient argument  to suppose that the energy - momentum of
 gravitational field of an attracting  mass is given by equation
(\ref{t=}) up to a term,
with the integral over the volume equal to zero.


\begin{thebibliography}{20}
\bibitem{Verozub1}  L.V.Verozub Phys.Lett. A 156 (1991) 404
\bibitem{Thirring} W.E.Thirring Ann.Phys. 16 (1961) 96
\bibitem{Verozub2}  L.V.Verozub Astron. Nachr. 317 (1996) 107
\bibitem{Abramovitz}  A.M.Abramowitz and I.A.Stegun Handbook of
Mathematical function , National of Buraw of Standards, 1964
\end{thebibliography}
\end{document}